\documentclass[pra,twocolumn,showpacs]{revtex4}
\usepackage{amsmath,amssymb,bm,txfonts,amssymb,epsfig,graphics}
\usepackage{CJK}

\begin{document}

\title{Processing Quantum Information in Hybrid Topological Qubit and
Superconducting Flux Qubit System}
\author{Zhen-Tao Zhang, Yang Yu}
\affiliation{National Laboratory of Solid State Microstructures, School of Physics,
Nanjing University , Nanjing 210093 , China}
\date{\today}

\begin{abstract}

A composite system of Majorana-hosted semiconductor nanowire and
superconducting flux qubit is investigated. It is found that the coupling
between these two subsystems can be controlled electrically, supplying a
convenient method to implement $\pi /8$ phase gate of a Majorana-based topological qubit. We also present a scheme
to transfer information from the flux qubit to the topological qubit using
Landau-Zener transition. In addition, a structure named top-flux-flux is
proposed to retrieve the information stored in the topological qubit. With the
demonstration of the entanglement of two topological qubits, it is very
promising to do quantum information process with this hybrid system.
\end{abstract}

\maketitle







\section{introduction}

Topological quantum computation in which information is encoded into
non-Abelian anyons is a promising approach to realize scalable quantum
computer. Topological qubits hold the merit of resistance to some local
fluctuations due to their non-local property. Recent progresses in the
physical realization of non-Abelian anyons of Ising type-Majorana fermion
(MF) have drew much attention in this field. It has been shown theoretically
that MF can exist as quasiparticles in many condensed matter systems,
including $p_{x}+ip_{y}$ superconductor \cite{Read00}, topological
insulator-superconductor heterostructures \cite{Fu08}, and
semiconductor-superconductor heterostructures \cite%
{Alicea10,Sau10,Oreg10,Lutchyn10}. Very recently, several groups have
reported the observation of MF zero energy mode at the ends of semiconductor
nanowire with the combination of spin-orbit coupling, proximity-induced
superconductivity and applied magnetic field \cite{Mourik12, Deng12,Das12}.
In addition, Alicea \textit{et al.} have shown that in the nanowire networks
MF, which obeys non-Abelian statistics, can be braided by
simply adjusting gate voltages. This system may furnish as a platform for
topological quantum computing.

For processing quantum information, two MF $\gamma _{1},\gamma
_{2}$ could be combined to form a Dirac fermion with creation and
annihilation operator as:
\begin{equation}
f=\frac{\gamma _{1}+i\gamma _{2}}{2},\indent f^{\dag }=\frac{\gamma
_{1}-i\gamma _{2}}{2}
\end{equation}%
The two states of the Dirac fermion, corresponding to $n=f^{\dag }f=0,1,$
could function as a qubit. Because braiding any two MF can not change the
parity of $n$, the same parity states of four MF are used to code one
topological qubit. For example, one typical choice is
\begin{equation}
|\psi \rangle =c_{1}|00\rangle +c_{2}|11\rangle .
\end{equation}

\indent However, by braiding MF one can not generate a complete set of
universal quantum logic gates required for quantum computation. It is
well known that single qubit $\pi/8$ phase gate and non trivial two-qubit gate
can not realized solely by braiding the object MF without auxiliary qubit. Therefore, we could not put whole
information process under topological protection. Moreover, for practical
quantum computation it is essential to transfer information between
topological qubits and other qubits. To fullfill these functions, people have
proposed many composite systems which consists of topological qubit and
conventional qubits, including superconducting flux qubit \cite{Sau10a,
Hassler10,Jiang11,Bonderson11}, transmon \cite{Hassler11}, quantum dot \cite%
{Bonderson11,Flensberg11,Leijnse11}. In this paper we investigate the hybrid
system of superconducting qubit and semiconductor nanowire which hosts MF.

It has been shown that flux qubit can be used to measure the state of
topological qubit using Aharonov-Casher effect [12]. One can also entangle
two qubits (topological qubit or double dot qubit or combination of them)
[11, 14] by jointly measuring them. Here we propose several important
schemes toward this direction. First, we show how to completely turn off the
coupling between flux qubit and topological qubit by setting appropriate
voltage on the capacitor. With this in hand, we can optimize the $\pi/8$
phase gate of topological qubit. Second, we employ Landau-Zener transition
to store the information of flux qubit into topological qubit. Third, We
present a scheme to retrieve the state of topological qubit back to a
well-coherence flux qubit. It is worth to emphasize that this is not a simple
inverse of the storage process. Because one flux qubit is occupied to
measure topological qubit, another flux qubit has to be used to receive the
information. Borrowing the technique of tunable coupling between flux qubits
[22], we have conceived a top-flux-flux composite system to realize the
information retrieval scheme. Our proposals build a viable interface between
topological and conventional qubits.

\bigskip

\section{tunable coupling between topological qubit and flux qubit and its
applications}

\subsection{method}

The hybrid system consists of a superconducting flux qubit coupled with a
nanowire which hosts some MF at its boundaries of
topological and un-topological parts (Fig 1). The flux qubit is made up of a
superconducting loop interrupted by three Josephson junctions. The junctions
have Josephson energy $E_{j},\alpha E_{j},E_{j}$, and charging energy $%
E_{c},\alpha E_{c},E_{c},$ respectively. In order to measure qubit and
transfer information, we choose $\alpha >1$ for optimizing Aharonov-Casher
effect. The Hamiltonian of the flux qubit is
\begin{equation}
H=-\frac{1}{2}(\varepsilon \sigma _{z}+\Delta \sigma _{x}),
\end{equation}%
where $\sigma _{z}$ and $\sigma _{x}$ are Pauli matrices. $\varepsilon
=2I_{p}(\phi -\frac{\phi _{0}}{2})$ is the magnetic energy of two diabatic
energy states $|L\rangle $ and $|R\rangle $, corresponding to the clockwise
and counterclockwise persistent current respectively. The magnetic energy is
adjustable by the external magnetic flux threading through the
superconducting loop $\phi$. $\phi _{0}$ is the single flux quantum. $I_{p}$
is the persistent current in the loop generated by $\phi $. $\Delta $ is the
tunneling splitting. At the energy level anti-crossing $\phi =\frac{\phi _{0}}{2}$ (so called optimal point), the two
energy states is degenerate and the tunneling coupling mixes them, resulting
a ground state $|g\rangle=(|L\rangle +$ $|R\rangle )/\sqrt{2}$ and a excite state $|e\rangle=(
|L\rangle -$ $|R\rangle )/\sqrt{2}$ with an energy splitting $\Delta $.
Because of Aharonov-Casher effect, the tunnel splitting is modulated by the
total charge on island and gate capacitor
\begin{equation}
\Delta =\Delta _{max}|\cos {(\pi q/2e)}|,
\end{equation}
where $q=en_{p}+q_{ext}$. $n_{p}$ = 0 (1) denotes the eigenvalue of the even
(odd) parity of the state of the fermion formed from the two MF located at the island. $q_{ext}$ is the total electrical charge in
the island and gate capacitor. If we calibrate the charge $q_{ext}$ to zero,
the splitting will be maximized when the fermionic state is in even parity
state. When the ferminon parity is odd, the splitting will be zero. In this
case, the energy of the two state is conditioned on the state of the
topological qubit. Therefore, we can measure the topological qubit state by
probing the tunneling splitting of the flux qubit.\newline
\indent However, the goal we are going to achieve using flux qubit is not only to
measure the state of topological qubit, but also to coherently transfer
information between them. Hence, it is necessary to switch on and off the
coupling efficiently and accurately. Our trick is to tune the voltage of
the capacitor to make sure that the both states of the topological qubit
lead to the same tunneling splitting of the flux qubit. In this case, the
eigenenergies of the flux qubit are not affected by the topological qubit,
which indicates that the two qubits are decoupled.
From Equation 4, it is easy to see that this can be realized by setting $q_{ext}=e/2$.
The value of the corresponding voltage added to the capacitor can be
obtained in the calibration step described above.
\begin{figure}[tbp]
\includegraphics [width=4cm]{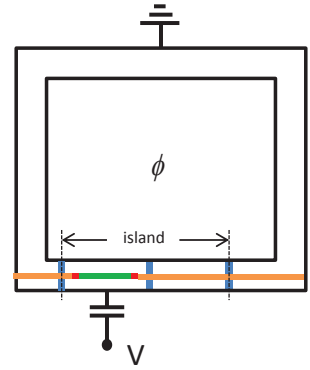}
\caption{(Color online) Flux qubit and semiconductor nanowire. The flux qubit consists of a superconducting loop interrupted by three Josephson junctions. The nanowire can be tuned to topological phase or non-topological phase in any sector. The boundary of these two phase will host MF (red points). The number of the charges on the island between the out two junctions affect the splitting of the flux qubit, i.e., Ahanaranov-Casher effect. We can control the number of the charges by tuning the voltage $V$ added to the island through a capacitor.}
\end{figure}
\subsection{implement $\protect\pi/8$ phase gate of topological qubit}

$\pi/8$ phase gate in topological qubit system can not be obtained by
braiding MF. We have devised a scheme to accomplish it with
the assistant of flux qubit. Firstly, the flux qubit is magnetically biased
away from $\phi =\phi _{0}/2$ and stay at its ground state. Before moving
the MF to the island, the electrical charge on the island and
the capacitor is tuned to $e/2$. Therefore, when the topological qubit is
introduced to the system, it is uncoupled to the flux qubit. Then we switch
on the coupling quickly. Depending on the state of the topological qubit,
the ground state have eigenenergy
\begin{equation}
E_{g}=\left\{
\begin{array}{cc}
-\frac{1}{2}\sqrt{\varepsilon +\Delta ^{2}},\indent & if\indent n_{p}=1 \\
-\frac{1}{2}\varepsilon ,\indent & if\indent n_{p}=0 \\
&
\end{array}%
\right.
\end{equation}

Therefore, the two states of the topological qubit coupled to the flux qubit
have a energy difference $\Delta E_{p}=\frac{1}{2}(\sqrt{\varepsilon
^{2}+\Delta ^{2}}-\varepsilon )$. We can realize any phase gate, including $%
\pi /8$ phase gate, by controlling the coupling time via tuning the
capacitor voltage.

\section{transfer information from flux qubit to topological qubit}
It is well known that transferring information between two qubits can be realized by using CNOT gate
combined with Hadamard gate and single-qubit measurement \cite{Nielsen}. Following this method, we have designed a scheme for
transferring information from flux qubit to topological qubit (Fig 2(a)).
Initially, the flux qubit is at an unknown pure state $|\psi \rangle
=a|g\rangle +b|e\rangle $, and the topological qubit stays at $|0\rangle $. Then apply
operations as following: Hadamard gate on the topological qubit, CNOT gate conditioned
on the topological qubit, and measurement on the flux qubit. If the measurement result is $%
|g\rangle $, the transferring process is successfully complete; otherwise,
an additional NOT operation should be applied to the topological qubit. \newline
\indent Let us turn to the question of how to implement each step. Actually,
Hadamard gate on topological qubit can be achieved by braiding MF in
one-dimension semiconductor nanowire network. The measurement of flux qubit
is in hand by now. However, the CNOT gate needs to be considered
deliberately. We have worked out a method which employs the coupling between
the two qubits and Landau-Zener transition.\newline
\indent Flux qubit is prepared in $|\psi \rangle $ at bias $\phi _{i}<\phi
_{0}/2$. The condition $\varepsilon >\Delta $ is required for the
Landau-Zener transition at the anti-crossing point well-defined. The
topological qubit stays at a superposition state $|\varphi \rangle
=(|0\rangle +|1\rangle )/\sqrt{2}$ after a Hadamard operation, and then is
loaded to the island. The flux qubit and topological qubit is uncoupled at
this moment. Then switch on the coupling by setting $q_{ext}=0$, and sweep
the flux bias through the anti-crossing point. If the topological qubit is $%
|0\rangle $ $(|1\rangle )$, the tunneling splitting is $\Delta _{max}$ $(0)$%
. Based on this feature, we can choose a sufficient low sweep velocity to
guarantee that the flux qubit evolutes adiabatically to the end without
destroying its state if the topological qubit state is $|0\rangle $, and
exchanges its ground state and excite state in the other case. Finally, turn
off the coupling and measure the flux qubit. It is clear that the sweeping
process is equivalent to a CNOT gate operation. \newline

Now, we estimate the minimum time to achieve the CNOT gate. In the
Landau-Zener transition formulism, the transition (between the ground state
and excite state) possibility is expressed as
\begin{equation}
\begin{array}{cc}
P_{1}=1,\indent & \indent n_{p}=1 \\
P_{0}=e^{-2\pi \Delta _{max}^{2}/4v},\indent & \indent n_{p}=0 \\
&
\end{array}%
\end{equation}%
where $v=\Delta \varepsilon /\Delta t$. Assuming the final energy bias is $%
-\varepsilon $, we have $v=2\varepsilon /\Delta t$. To make $P_{0}=e^{-2\pi
\Delta _{max}^{2}/4v}=e^{-2\pi \Delta _{max}^{2}\Delta t/8\varepsilon
}\approx 0$, we get $2\pi \Delta _{max}^{2}\Delta t/8\varepsilon \gg 1$.
If $\varepsilon =2\Delta _{max}=2\times 2\pi $ $GHz$, $\Delta t$ should
much larger than $0.4$ $ns$. We could choose $\Delta t=10$ $ns$, which is
much shorter than the coherence time of the flux qubit.
\begin{figure}[tbp]
\includegraphics [width=7cm]{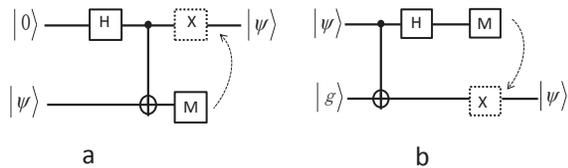}
\caption{Quantum circuit for quantum state transfer between
topological qubit and flux qubit. (a)Transfer a unknown state $|\protect\psi %
\rangle $ from the flux qubit to the topological qubit. (b)Transfer a
unknown state $|\protect\psi \rangle $ from the topological qubit to the
flux qubit.}
\end{figure}

\section{transfer information from topological qubit to flux qubit}
In the previous section, we have addressed the question of how to "write"
the data of the flux qubit to the topological qubit. Similarly, "read" the
data of the topological qubit, i.e, transferring information from
topological qubit to flux qubit, is also important for the hybrid system. In
principle, it can be done by constructing a similar process like that in the last section. The initial state of the
topological qubit and flux qubit is $|\psi \rangle =a|0\rangle +b|1\rangle $
and $|g\rangle $ respectively. Apply in order the operations: CNOT gate, Hadamard gate,
and measurement on the topological qubit [Fig 2(b)]. At last,
add a NOT (Identity) operation to the flux qubit if the readout result is $%
|1\rangle $ $(|0\rangle )$. It seems straightforward to achieve in
experiment, because we can use directly the relevant techniques illustrated
in the previous section. However, attention must be paid to the differences
between them: \newline

\indent1. After the CNOT operation, the topological qubit is subject to Hadamard operation and
measurement. Meanwhile, the flux qubit needs to keep coherent. This require a longer coherence
time of the flux qubit compared to that in the former section.
\newline
\begin{figure}[tbp]
\includegraphics [width=5.5cm,height=3.5cm]{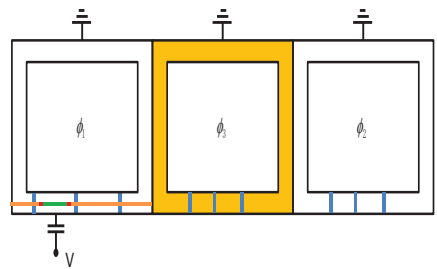}
\caption{(Color online) Sketch of top-flux-flux. The two flux qubits are
coupled through the loop between them. When a pulse of microwave with
frequency equal to the difference of the two flux qubits is added to the
mediate loop, the coupling is on. When the microwave is turned off, the
coupling is off.}
\end{figure}
\indent2. The object of the measurement is not the flux qubit but the
topological qubit. Therefore one flux qubit is not enough here. \newline
\indent In order to make the information transferring feasible, we
have developed a new setup (see Fig 3), which is dubbed top-flux-flux structure.
In this setup, two flux qubits (named qubit 1 and 2) are used. Qubit 1 is
just the same as that introduced in the previous section, and functions as
readout device of the topological qubit and information medium between
topological qubit and qubit 2. Qubit 2 is a conventional
three-junction flux qubit with $\alpha <1$. Its eigenergies are insensitive
to charge fluctuation due to the absence of the Aharonov-Casher effect. Hence, its coherence
time could be longer than qubit 1, which makes it more suitable
as an information receiver.\newline
\indent Before showing the information transfer procedure, we briefly describe how the two flux
qubits couple. Generally, two flux qubits can interact directly through
geometric mutual inductance \cite{Majer05}. However, the generated
interaction is not prone to be switch off, which makes individual qubit
operation unrealistic. Instead, we adopt the scheme proposed and
demonstrated experimentally by Niskanen et al.\cite{Niskanen06,Niskanen07}.
Qubit 1 and qubit 2 are coupled through a third qubit between them. When both
qubits are biased at the optimal point with splitting $\Delta _{1},\Delta
_{2}$ respectively, applying a microwave with frequency $\omega =|\Delta
_{1}-\Delta _{2}|$ to the coupler qubit turns on the interaction with form
in rotating frame:
\begin{equation}
H_{i}=\Omega (\sigma _{x}^{1}\sigma _{x}^{2}+\sigma _{y}^{1}\sigma _{y}^{2})
\end{equation}%
Where $\Omega $ is the oscillation frequency between $|ge\rangle $ and $%
|eg\rangle $. The interaction is off when the microwave is off.

\indent Now we explain our information transferring protocol in detail.
Qubit 1 is prepared at ground state and biased at point A (Fig 4), which is far away
from the optimal point;
 qubit 2 is prepared at its excite state and biased at the optimal point.
One may ask why qubit 2 is not prepared at the ground state as addressed at the
beginning of this section. Actually, the purpose of preparing qubit 2 at
the ground state and performing a CNOT operation on qubit 2 and the
topological qubit is to produce the entangle state $(a|0g\rangle
+b|1e\rangle )$ of them. We will demonstrate in the following that
the entanglement state can also be generated with qubit 2 prepared at its excite state.
Initial state of qubit 1 is approximately $|L\rangle $. Ever since the
MF are loaded into the island of the qubit 1, turn on the
coupling between topological qubit and qubit 1. Then, we sweep the bias of
qubit 1 adiabatically to the anti-crossing point. As a result, the state of
qubit 1 will remain at the ground state if the topological qubit is $|0\rangle$
because the splitting at the anti-crossing point is maximized. If
the topological qubit is $|1\rangle $, qubit 1 will stay at its initial
state $|L\rangle $ without mixing with $|R\rangle $. Then we add the
microwave with frequency $\omega =|\Delta _{1}-\Delta _{2}|$ to the coupler
to switch on the interaction between qubit 1 and 2. Due to the resonance
condition the flux qubits interact only if the topological qubit is $%
|0\rangle $. Choosing the microwave pulse with duration $1/2\Omega $, we
have
\begin{equation}
(a|0g\rangle +b|1L\rangle )|e\rangle \rightarrow a|0eg\rangle +b|1Le\rangle
\end{equation}%
Now the topological qubit is entangled with two flux qubits. The next
step is separating the qubit 1 from the entanglement. This is achieved by
adiabatically sweeping the external flux bias of the qubit 1 across the
anti-crossing point to point B (see Fig 4) which is far away from the anti-crossing. At
the end, the state $|L\rangle $ of qubit 1 is equal to $|e\rangle $.
Therefore the final state of topological qubit and qubit 2 is $(a|0g\rangle
+b|1e\rangle )$. The remaining operations are straightforward: braiding the MF
to realize Hadamard gate, measure the topological qubit with qubit 1, and so
on.

\begin{figure}[tbp]
\includegraphics [width=4.3cm]{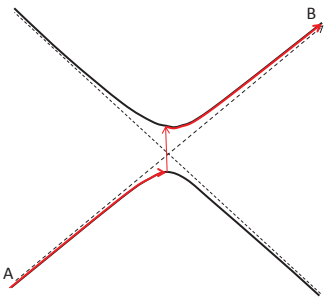}
\caption{(Color online) Energy diagram and the evolution of flux qubit 1 for
quantum information transferring from flux qubit to topological qubit. The flux
qubit 1 is prepared at ground state and biased at point A. Sweep the bias adiabatically to the anti-crossing point.
Companied with different state of the topological qubit, flux 1 evolute
along solid (red) line when topological qubit is $|0\rangle $, along the
dashed line when topological qubit is $|1\rangle $. After shortly coupling
with qubit 2 at the optimal point, bias is swept adiabatically to
right until the point B, where qubit 1 stay at excite state for both states
of the topological qubit.}
\end{figure}
It is worth to note that adiabatic condition is needed in the bias
sweeping process for qubit 1. The adiabatic condition can be
characterized by Landau-Zener transition possibility, and is satisfied
if the transition probability in the sweeping process is vanishing. The Landau-Zener transition probability is
\begin{equation}
P_{LZ}=e^{-2\pi\Delta_1^2/4v}
\end{equation}
where $\Delta _{1}$ is the energy splitting at the optimal point when
topological qubit is at $|0\rangle $, $v=2\varepsilon /\Delta t$, $%
\varepsilon $ is the bias energy at the initial bias, $\Delta t$ is the time of the sweeping. The adiabatic
condition $P_{LZ}\approx 0$ is met when $e^{-2\pi \Delta _{1}^{2}/4v}\approx
0 $. Assuming $\varepsilon =10\Delta _{1}=20\times 2\pi GHz$, the time scale of this
process should be much longer than $1ns$. It is sufficient if we set the duration of the
sweeping process $\Delta t=20ns$. Besides, the coupling time of
qubit 1 and 2 scales as $1/2\Omega $ which is typically $\sim 20ns$
\cite{Niskanen07}. In all, the generation of the entanglement between the
topological qubit and qubit 2 can be done within $\sim 40ns$, which is much
shorter than the coherence time($\sim 2\mu s$) of qubit 2.
\section{discussion}
\indent  We have noticed that implementing $\pi/8$ phase gate of topological qubit with the help of flux qubit was considered by Hassler \emph{et al}. before \cite{Hassler10}. They also employ the energy difference $\Delta E_g$ of the ground state of the flux qubit resulting from the parities of different topological states to accumulate phase difference. Finally, they  decouple the two qubits by biasing the external flux of flux qubit far away from the optimal point. However, this method can not turn off the coupling completely because even when we bias flux far from the optimal point to make $\varepsilon\gg\Delta$, the energy difference $\Delta E_g$ is non-vanishing up to the first order of $\Delta/\varepsilon$. This brings error to the $\pi/8$ phase gate. Our scheme gets rid of this problem because we can turn off the coupling completely by electrical control. People also suggested using transmon \cite{Koch07}, a variant of superconducting charge qubit, to measure and communicate with topological qubit \cite{Hassler11}. One can simply consider transmon as two superconductor islands connected by a dc SQUID. The advantage of using it is that the coupling with topological qubit can be switched on and off with exponential accuracy \cite{Hassler11}. However, there are two drawbacks in their measurement scheme: 1. For measuring single topological qubit, all four MF consisting of the qubit should first be moved to one island of a transmon, then two of them are transfer to the other island. The process is much more complex than that of using flux qubit, and the situation is even worse when doing joint measurement. 2. The state calibration is very challenging. If the capacitances of the two islands are not symmetry, the period of the charge vs the frequency of the transmon is not $2e$ (see appendix of Ref \cite{Hassler11} for detail). Therefore when the four MF are loaded into one island of the transmon, the two constituent state $|00\rangle$ and $|11\rangle$ correspond to different frequencies of the transmon, which makes the calibration of the charge on the islands impossible. Hence, the asymmetry of the the islands will reduce the practicability of transmon as a topological qubit measurer. On the contrary, flux qubits are not bothered by this kind of asymmetry. From this point of view, flux qubit might be a better candidate for measuring topological qubit. That is why we choose it as a interface between topological and conventional qubit system.

\section{conclusion}

We have present a method to electrically control the coupling between
topological qubits and flux qubits from which we can implement $\pi/8$ phase
gate of a topological qubit. Combined with generating entanglement through joint measurement of two
topological qubits and the braiding operations of MF, the hybrid system of
semiconductor nanowire and flux qubit possesses a set of universal quantum
logic gates for realizing universal quantum computation. We also propose a scheme to transfer
information from a flux qubit to a topological qubit via Landau-Zener transition.
In addition, we have conceived a structure of top-flux-flux to retrieve the
information stored in a topological qubit. Therefore, we have constructed a
viable interface between topological qubit system and conventional quantum
system.\\

This work was supported in part by MOST of China (2011CB922104, 2011CBA00200), NSFC (91021003, 11274156), the Natural Science Foundation of Jiangsu Province (BK2010012), and PAPD.


\end{document}